\documentclass[aps,pra,10pt,twocolumn,superscriptaddress]{revtex4}
\usepackage{amsmath}
\usepackage{amssymb}
\usepackage{bbm}
\usepackage{graphicx}
\usepackage{framed}
\usepackage{color}
\usepackage{hyperref}
\usepackage{epstopdf}
\usepackage{dsfont}
\usepackage{amssymb}
\usepackage{amsmath}
\usepackage{amsthm}
\usepackage{wrapfig}
\usepackage{relsize}
\usepackage{bm}
\usepackage[latin1]{inputenc}
\usepackage{enumitem}

\newlist{myitemize}{enumerate}{10}
\setlist[myitemize]{label*=\arabic*.,nosep,leftmargin=*}

\newcommand{\ba}{\begin{eqnarray}}
\newcommand{\ea}{\end{eqnarray}}
\newcommand{\pd}[2]{\frac{\partial #1}{\partial #2}}

\newcommand{\identity}{\mathlarger{
\mathds{1}}}
\newcommand{\bra}[1]{ \langle #1  |}
\newcommand{\ket}[1]{ | #1 \rangle }
\newcommand{\bracket}[2]{{\langle #1}|{#2 \rangle}}

\begin{document}

\title{A variational approach to the optimal control of coherently driven, open quantum system dynamics}

\author{Vasco Cavina}
\affiliation{NEST, Scuola Normale Superiore and Istituto Nanoscienze-CNR, I-56126 Pisa, Italy}
\author{Andrea Mari}
\affiliation{NEST, Scuola Normale Superiore and Istituto Nanoscienze-CNR, I-56126 Pisa, Italy}
\author{Alberto Carlini} 
\affiliation{Università degli Studi del Piemonte Orientale Amedeo Avogadro, Italy}
\author{Vittorio Giovannetti}
\affiliation{NEST, Scuola Normale Superiore and Istituto Nanoscienze-CNR, I-56126 Pisa, Italy}

\begin{abstract} 

Quantum coherence inherently affects the dynamics and the performances of a quantum machine.
Coherent control can, at least in principle, enhance the work extraction and boost
the velocity of evolution in an open quantum system.
Using advanced tools from the calculus of variations and reformulating the control problem in the instantaneous Hamiltonian eigenframe,
we develop a general technique for minimizing
a wide class of cost functionals when the external control has access to full rotations of the 
system Hamiltonian. 
The method is then applied both to time and heat loss minimization problems and explicitly solved 
in the case of a two level system in contact with either bosonic or fermionic thermal environments.

%

\end{abstract}

\maketitle

\section{Introduction}

%
%

Differently from the universal results of classical thermodynamics such as the first and the second law, the analysis of quantum systems driven out-of-equilibrium involves non universal features depending on the details of the dynamics \cite{Kosloff2013, Cavina2017, Avron2012} or on the response of the system to an external perturbation \cite{Benenti2017, Bonanca2014, Davies1978}. In such irreversible situations, 
optimizing thermodynamic quantities like heat or work usually requires non-trivial control strategies 
 that explicilty involve quantum operations \cite{stefanatos2014, campaioli2017, zheng2016, deng2018}.
In this framework optimal control theory has proved to be effective for solving a variety of applicative tasks \cite{bathaee2016, Sauer2013, Ma2018, Einax2014, Wang2012}. 
Beyond thermodynamics, optimal control theory is well known to be useful in time minimization problems \cite{Lapert2013, Doria2011}, for the study of quantum speed limits \cite{Caneva2009} and for generating efficient quantum gates in dissipative systems \cite{Bonnard2009,Roloff2009,Schulte2011, koch2016}.
Such different goals can be achieved with 
several techniques, depending on the framework in consideration, i.e. on the dynamical equations and the physical constraints associated with the controlled system, and on the quantites one wants to optimize.
These methods span, e.g., from Floquet theory, that is particularly suitable for periodical external driving forces \cite{Sigmund1997,Guerin2007}, to geometric reformulations of the control problem \cite{Khaneja2001, Boscain2001,Khaneja2003,Sugny2007,Bonnard2009,Yuan2012,Zhang2015,Rotskoff2017} 
for fidelity or time optimization and applications in the linear response
regime, to adiabatic and shortcut to adiabaticity approaches \cite{Sarandy2005,Joye2007}. An excellent review on the recent advances in optimal quantum control theory can be found in \cite{Glaser2015}.

In this work we focus on externally driven open quantum systems and we develop a formal variational approach which is general enough to cover thermodynamics and time minimization problems. 
We will use a powerful tool known as the Pontryagin Mimimum Principle (PMP) \cite{Kirk2004}, already succesfully applied in time \cite{Carlini2006, suri2017, Stefanatos2013, Mintert2011, Assemat2010} and heat loss \cite{Cavina2017bis} optimization problems. The peculiarity or our work is that we consider quantum systems which are open ({\it i.e.}\ in contact with a thermal bath) and which might develop quantum coherence between the energy eigenstates. The latter is an intrinsically quantum mechanical effect which is often neglected in many thermodynamic analysis but which, at least in principle, could allow for better optimization strategies with respect to a semi-classical driving of the system.

For this sake we will suppose that the dynamics of the system weakly coupled to a thermal bath is described by a Markovian master equation (MME) of the Lindblad form \cite{Lindblad1976, Gorini1976}
\ba \label{dyn}  
\frac{d \rho(t)}{dt} = \mathcal{L}_{\bold{u}(t)}[{{\rho}}(t)] := - i [{H}_{\bold{u}(t)}, {\rho}(t)] + \mathcal{D}_{\bold{u}(t)}[{{\rho}}(t)]\;, 
\ea
where ${H}_{{\bf u}(t)}$ is the system Hamiltonian and $\mathcal{D}_{{\bf u}(t)}$ is the Gorini-Kossakowski-Sudarshan-Lindblad (GKSL) dissipator,
and both are assumed to implicitly depend on a family of external control fields that we cast in the form of the vector ${\bf u}(t) = [u_1(t),u_2(t),...]$ (throughout the paper we will use the convention that $\hbar =1$). 
We are interested in the problem of minimizing general cost functions associated to the state evolution of the system from an initial time $t=0$ to a final time $t=\tau$, and possessing the following structure
\begin{figure}[t]
\includegraphics[width=0.5 \textwidth]{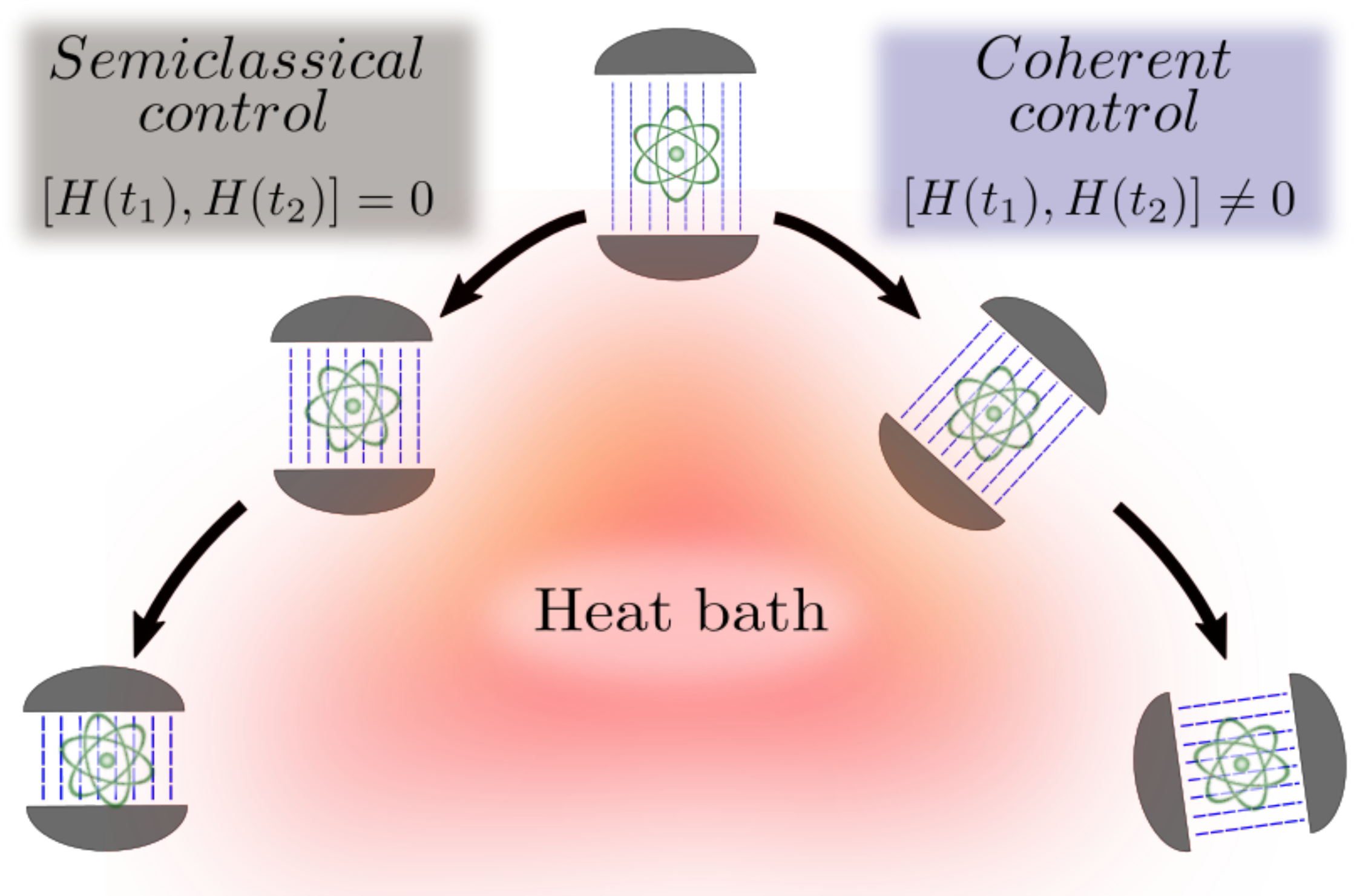}
\caption{Pictorial representation of two possible strategies to control a quantum system in the time interval $[0,\tau]$: on the left the 
eigenvectors of the Hamiltonian are fixed i.e. $[H(t_1), H(t_2)]=0$; on the
right the Hamiltonian can rotate and $[H(t_1), H(t_2)]\neq 0$. In the following we will provide a set of 
necessary coonditions for an optimal control specifically in this last case.} \label{fig:1}
\end{figure}

\ba \label{generalf} 
f :=  \int_0^{\tau} 
\Big\langle \mathcal F_{\bold{u}(t)} [{{\rho}}(t)] \Big\rangle dt \;, 
\ea
where $\mathcal F_{\bold{u}(t)}$ is a generic control-dependent linear operator acting on the quantum state, while the brackets $\langle \cdot \rangle$  denote the trace operation. 

In this work we aim principally at the development of a formalism for handling quantum coherences in the variational calculus.
In Section \ref{variational} we show that such a problem can be tackled 
by doing a time dependent change of basis that brings the system into the instantaneous Hamiltonian eigenframe and 
by introducing a convenient reparameterization of the control fields.

In Section \ref{sec:heat} we apply this formalism to the heat minimization problem and we present three physical models as examples: a two-level system in a Gibbs mixing channel and a two-level system in a thermal bath with either
bosonic or fermionic excitations.
Eventually in Section \ref{Aqs} we will see that our general approach proves to be useful also for solving time minimization problems and for characterizing the set of reachable states for open quantum systems. In order to make the main text easier to read, we moved many details and calculations in technical appendices. 

\section{General variational approach in a rotating frame}
\label{variational}
A stationary solution of the functional (\ref{generalf}) under the constraint (\ref{dyn}) can be found through an extremization of the extended functional 
\ba\label{eqfNew}
{\cal{J}}&:=&f + \int_0^{\tau} \;  \Big\{ \lambda(t) ( \langle{\rho}(t) \rangle- 1)
\\ 
&&\qquad +\Big\langle \mathcal{\pi}(t) \left ( \mathcal{L}_{\bold{u}(t)}[{{\rho}}(t)]-\frac{d \rho (t)}{dt} \right )\Big\rangle\Big\}\;dt  \nonumber \;,
\ea
where $\pi(t)$ is a self-adjoint traceless \cite{Nota0} operator and $\lambda(t)$ is a scalar, respectively acting 
as Lagrange multipliers of the dynamical constraint (\ref{dyn}) and of the normalization of the state $\rho$. 
Notice that all the variables appearing in  Eq. (\ref{eqfNew}) are independent, 
thus the integrand on the right hand side is {\it a priori} different from zero, although it nullifies {\it on-shell} as 
a consequence of the minimum conditions \cite{Kirk2004}.    
The functional (\ref{eqfNew}) is the starting point of the PMP approach, cf. for instance Ref. \cite{Cavina2017bis}, in which 
this kind of functional was introduced to study heat loss optimization problems.
Let us suppose now that the system Hamiltonian $H_{{\bf u}(t)}$ is fully controllable, {\it i.e.}
the external control fields can be tuned to obtain a generic
self-adjoint Hamiltonian with time-dependent eigenvectors and eigenvalues (see Fig. \ref{fig:1}). 
It is convenient to parameterize the Hamiltonian through its spectral decomposition
\ba \label{rot} 
H_{{\bf u}(t)}= U^{\dagger}(t)D(t)U(t),   
\ea
where $U(t)$ and $D(t)$ are respectively a unitary matrix containing the eigenvectors and a real diagonal matrix containing the energy levels.
These two objects are just a different parameterization of the control fields and so, from now on, we drop the subscript ${\bf u}(t)$ for ease of notation.

A semi-classical modulation of the energy levels corresponds to keeping $U(t)$ equal to the identity and this regime has been often studied in the context of quantum thermodynamic processes (see, e.g., Refs. \cite{Cavina2017bis,Esposito2010}). Quantum mechanics however allows for a larger class of possible controls where, in addition to the manipulation of the energy eigenvalues $D(t)$, also the energy eigenstates can be rotated by a non-trivial unitary matrix $U(t)$. The main task of this work is to develop a formalism which is suitable also for this coherent regime. The idea is to introduce a reference frame which is co-moving with the Hamiltonian in such a way that, in the rotating frame, $H(t)$ always looks like a semiclassical diagonal matrix. The corresponding quantum state and co-state in this frame are given by
\ba
 \tilde{\rho} (t) &=& U(t) \rho(t) U^{\dagger}(t), \label{rhoTilde}\\
 \tilde{\mathcal{\pi}} (t) &=& U(t) \mathcal{\pi}(t) U^{\dagger}(t).\label{piTilde}
\ea

\noindent 
Moreover, we can express $U(t)$ in terms of a self-adjoint operator $\Lambda(t)$, in such a way that the motion of the rotating frame is represented as induced by a fictitious Hamiltonian term $\Lambda(t)$.   
If $U(t)$ is sufficiently regular, {\it i.e.} its entries are continuous and differentiable, it is possible to cast it in terms of a time ordered exponential
\ba \label{gen} 
U(t)= \overrightarrow{\exp}\Big( \int_0^t i \Lambda(t') dt'  \Big)  U(0), 
\ea
which is the solution of the following differential equation
\ba \label{diffrot}
 \dot{U}(t) =  i \Lambda(t) U(t), 
 \ea
with initial condition $U(0)$. According to Eq. (\ref{gen}), $\Lambda(t)$ is the generator of the change of basis which diagonalizes the Hamiltonian.
Moreover it is easy to check that the time derivative of the quantum state satisfies
\ba  \label{rotdin} 
U(t) \dot{\rho}(t) U^{\dagger}(t) = \dot{\tilde{\rho}}(t) - i[\Lambda(t), \tilde{\rho}(t)]. 
\ea

Now we make an important assumption about the structure of the generic functional $\mathcal F_{\bold{u}(t)}$ introduced in Eq.\ \eqref{generalf} which, as we are going to show, applies to many  practical situations.\\
 
\noindent \textbf{Assumption 1 ($H$-covariance):} We assume that $\mathcal F_{\bold{u}(t)}$ may depend non-trivially only on the energy levels $D(t)$ of the Hamiltonian $H_{{\bf u}(t)}$, while it is covariant with respect to Hamiltonian rotations, {\it i.e.}
 \ba \label{Frot} 
\mathcal{F}_{u(t)}[\rho(t)] = U^{\dagger}(t)\mathcal{F}_{D(t)}[\tilde\rho(t)] U(t), 
\ea
where $U(t)$ and $D(t)$ are the matrices defined in \eqref{rot}. In what follows we denote all linear operators which obey the previous property as {\it$H$-covariant}. \\

\noindent Simple examples of $H$-covariant operators are the left and right multiplications of $\rho(t)$ by $H(t)$ or any analytical function of $H(t)$.
Another important example is given by the class of thermal Liouvillian operators, {\it i.e.}\ the class of generators of the thermal master equation introduced in Eq.\eqref{dyn}. Indeed, by following the standard microscopic interpretation of Eq.\ \eqref{dyn} as an effective map emerging from the interaction of the system with a heat bath, one can easily show (see Appendix \ref{app:dissipator}), that thermal dissipators and Liouvillian operators are H-covariant. More explicitly,

\ba \label{Drot} 
\mathcal{L}_{u(t)}[\rho(t)] = U^{\dagger}(t) \mathcal{L}_{D(t)}[\tilde\rho(t)] U(t)  \ea
where   
\ba \label{dynD}  
\mathcal{L}_{D(t)}[{\tilde{\rho}}(t)] := - i [D(t), \tilde{\rho}(t)] + \mathcal{D}_{D(t)}[{\tilde{\rho}}(t)]\;, 
\ea
and $\mathcal{D}_{D(t)}$ is the GKSL thermal dissipator associated to the diagonal Hamiltonian $D(t)$.

With this in mind, it is possible to rewrite the extended functional (\ref{eqfNew}) in terms of the rotated variables \eqref{rhoTilde} and \eqref{piTilde}. Making use of Eq.\ \eqref{rotdin} we obtain
\ba\label{extfunctional1New} 
\begin{gathered}
{\cal{J}}=  \int_0^{\tau} \Big\{\lambda(t) ( \langle\tilde{\rho}(t) \rangle- 1)+ \Big\langle  \mathcal F_{D(t)} [{\tilde{\rho}}(t)] \\
+ \mathcal{\tilde{\pi}}(t) \left ( \mathcal{L}_{D(t)}[{\tilde{\rho}}(t)]-\frac{d {\tilde{\rho}}(t) }{dt} + i[\Lambda(t),\tilde{\rho}(t)]\right )
\Big\rangle \Big\}\;dt. \end{gathered}
\ea
At first glance our choice to parameterize the system in terms of the transformed
variables $\tilde{\pi}(t)$, $\tilde{\rho}(t)$, $D(t)$ and the generator $\Lambda(t)$ may seem quite arbitrary and unnecessarily contrived.
However the great advantage in doing such an operation is that the extended functional \eqref{extfunctional1New} is now {\it linear} in $\Lambda(t)$ which allows to significantly simplify the problem.

In fact, following the standard approach used in classical control theory \cite{Kirk2004}, we first map the Lagrangian minimization problem \eqref{extfunctional1New} into the so called pseudo-Hamiltonian and then we apply the PMP. Thus, the functional $\cal J$ can be rewritten as
\ba\label{extfunctional}
{\cal{J}}=\int_0^{\tau} \; \Big\{ {\mathcal{H}}(t) -\Big\langle \tilde{\mathcal{\pi}}(t)\frac{d {\tilde{\rho}}(t)}{dt} \Big\rangle \Big\} \; dt,\;\label{lagr}
\ea
where 
\begin{eqnarray} 
{\mathcal{H}}(t):=\Big\langle \big(\tilde{\mathcal{\pi}}(t) \mathcal{L}_{D(t)}[\tilde{\rho}(t)]+\mathcal F_{D(t)} [{\hat{\rho}}(t)] \Big\rangle \\
+ \lambda(t) ( \langle \tilde{\rho}(t)\rangle - 1) + i\langle \Lambda [\tilde{\rho}(t), \tilde{\pi}(t)]\rangle  \nonumber  \label{HHNew} 
 \end{eqnarray} 
is the pseudo Hamiltonian. It is important to remark that $\mathcal H(t)$ is just a mathematical object associated with the control problem and it is completely different from the physical Hamiltonian $H(t)$ of the quantum system. Now we can finally apply the PMP \cite{Kirk2004} which establishes three necessary conditions that have to be satisfied by all extremal solutions of the extended functional. The first condition states that {\it i)} a non-zero costate $\tilde{\pi}(t)$ exists such that the following pseudo Hamilton equations hold
 \begin{eqnarray} 
 \frac{d{\tilde{\rho}(t)}}{dt} = \pd{\mathcal{H}(t)}{{\tilde{\pi}(t)}} \;,  \qquad 
 \frac{d{\tilde{\pi}}(t)}{dt} = - \pd{\mathcal{H}(t)}{\tilde{\rho}(t)}\;.   \label{H1}
 \end{eqnarray} 
The previous equations of motion determine, in the rotating frame identified by $U(t)$, the  dynamical evolution of the state and of the costate.
The second condition states that {\it ii)}
 for all
 $t \in [0,\tau]$ the pseudo Hamiltonian $\mathcal{H}(t)$ 
 has to be a minimum with respect to the control fields, that in our case are the entries of $\Lambda(t)$ and $D(t)$, and {\it iii)} it 
must assume a constant value $\mathcal{K}$, i.e.
 \ba \label{cons} 
\mathcal{H}(t)= \mathcal{K}.
 \ea

The minima of the functional (\ref{eqfNew}) subject to the dynamical constraint (\ref{dyn}) are obtained by imposing the previous prescriptions as described in more details in Appendix \ref{app:heat-general} (see also Ref. \cite{Cavina2017bis} for a similar treatment).

The same approach is valid both for fixed or free initial and final states, but we recall that the boundary conditions 
are functions of the original state variable $\rho(t)$, and not of its rotated version $\tilde{\rho}(t)$.
Thus, from Eqs. (\ref{rhoTilde}) and (\ref{gen}) we have that
$\tilde{\rho}(0) = U(0) \rho(0) U^{\dagger}(0)$ and $\tilde{\rho}(\tau) = \overrightarrow{\exp}\Big(i \int_0^{\tau} \Lambda dt \Big) U(0) \rho(\tau) U^{\dagger}(0) \overrightarrow{\exp}\Big(-i\int_0^{\tau} \Lambda dt \Big)$.
Finally we stress again that, in order to obtain the previous relations, we are assuming that the control fields
are sufficiently smooth.
If we broaden our analysis allowing piecewise smooth solutions we have to impose the so-called Weierstrass-Erdmann conditions 
stating the continuity of $\tilde{\pi}(t)$ and $\mathcal{H}(t)$ at the corner points  \cite{Kirk2004}.
The controls can be discontinuous at these points,
while $\tilde{\rho}$ can undergo an instantaneous unitary rotation, obtained, for instance, applying a divergent 
Hamiltonian for an infinitesimal period of time.
These irregular trajectories are an idealized mathematical limit of an extremely fast and effectively adiabatic process,
that in practice can occur when the external fields are varied on a time scale much smaller than those typically 
emerging from the naked (i.e., without controls) dissipative dynamics. 
Examples of such kind of control strategies in open quantum systems have been theoretically considered in Refs. \cite{Cavina2017bis, mukherjee2013}, while experimental implementations have been realized, for example, using electron islands \cite{Koski2014}.

The general approach presented in this Section applies to the minimization of a generic cost function \eqref{generalf} determined by an arbitrary, $H$-covariant,  linear operator $\mathcal F_{\bold{u}(t)} [{\hat{\rho}}(t)]$.  In the next Sections we are going to consider some relevant applications in different contexts, i.e. quantum thermodynamics and quantum speed-limits.

\section{Minimization of heat dissipation in coherent systems}
\label{sec:heat}
Given the dynamical evolution of an open quantum system according to the thermal master equation \eqref{dyn}, the amount of heat dissipated by the system into the environment in a time $\tau$ is given by  \cite{Alicki1979, Anders2013, Kosloff2013}
\ba \label{heat} 
Q := - \int_0^{\tau} 
\Big\langle \hat{H}_{\bold{u}(t)} \; \mathcal{L}_{\bold{u}(t)}  [{\hat{\rho}}(t)] \Big\rangle dt \;.
\ea
In the semi-classical case, {\it i.e.} when the state, the costate and the Hamiltonian remain diagonal, the optimal control problem for minimizing heat dissipation has been already studied \cite{Esposito2010,Cavina2017bis}. Here, our aim is to consider the larger set of possible control strategies in which quantum coherences can be created during the time evolution. 
For this task, we use the formalism developed in the previous Section and replace the general linear operator in Eq. \eqref{generalf} with the heat flux operator
\ba
F_{\bold{u}(t)} [{\hat{\rho}}(t)]=-\hat{H}_{\bold{u}(t)} \; \mathcal{L}_{\bold{u}(t)}  [{\hat{\rho}}(t)],
\ea
so that the generic cost function $f$ in Eq. \eqref{generalf} becomes equal to the dissipated heat $Q$ defined in Eq. \eqref{heat}. 

In this particular case, in addition to the dynamical equations \eqref{H1} and to the conserved quantity \eqref{cons} originating from the PMP, we can perform further algebraic manipulations (see Appendix \ref{app:heat-general}) obtaining the following additional relations 
\begin{gather} \label{3} 
[\tilde{\mathcal{\pi}}(t), \tilde{\rho}(t)] = 0; \\ \,
\label{4} [\tilde{\mathcal{\pi}}(t), \mathcal{L}_{D(t)}[\tilde{\rho}(t)]] + [\tilde{\rho}(t), \mathcal{L}_{D(t)}^{\dagger}[\tilde{\mathcal{\pi}}(t)]] =[\tilde{\rho}(t),\mathcal{L}_{D(t)}^{\dagger}[ D(t)]].
\end{gather}
The previous conditions are particularly appealing because they are simple matricial algebraic equations. In particular, despite  Eq. (\ref{4}) can be obtained from Eqs. (\ref{H1}, \ref{3}) thus being redundant in the PMP set of solutions,
 it is nevertheless very useful since we can trade it with one of the more difficult differential equations (\ref{H1}).
In the following we will apply the formalism developed above to two specific models of dissipation described by a MME in the Lindblad form
(\ref{dyn}). For this reason, although we are considering unconstrained families of Hamiltonians, we have to ensure that the driving is sufficiently 
slow and the energy gaps of the $D$ matrix are sufficiently large in order to preserve the Born-Markov and the secular approximations \cite{Breuer2002}.
If the optimal control history does not fullfill these conditions we have to introduce non Markovian corrections to Eq. (\ref{dyn}) in order to 
get a more physical and realistic description.

\subsection{Two-level system in a Gibbs mixing channel}
\label{Cpm}
As an example of coherent optimization we consider a two level system evolving through a master equation (\ref{dyn}) with a dissipator of the form
\begin{equation} \label{master}  
\mathcal{D}_G
[ \hat{\rho}(t)]  = \gamma[\hat{\eta}_{\beta}(t) - \hat{\rho}(t)], 
\end{equation}
where $\hat{\eta}_{\beta}(t)$ is the Gibbs state associated with the Hamiltonian $H_{\bf u}(t)$ and the inverse temperature $\beta$, while $\gamma$ is the decoherence rate.
For this model, the optimal trajectories minimizing the functional (\ref{heat}) are known only for semi-classical processes \cite{Cavina2017bis, Esposito2010} while
the formalism introduced in the previous Section paves the way to a general discussion.
After the change of basis (\ref{rot}) the Hamiltonian $D(t)$ will be a linear combination of  $\identity$
and $\sigma_z$ but, since the term proportional to the identity is arbitrary \cite{notegauge}, we can always set the ground state energy to zero such that  
\ba \label{hameps}  
D(t) = \frac{\epsilon(t)}{2} (\identity + \sigma_z), 
\ea
where $\epsilon(t)$ is the energy of the excited state.
The state and the costate can be parameterized using a pair of Bloch vectors $\vec{a}(t)$ and $\vec{q}(t)$, {\it i.e.}
\ba
\rho(t) &=&\frac{1}{2} [\identity + \vec{a}(t) \cdot \vec{\sigma}] \label{rhoBloch} \\
\mathcal{\pi}(t) &=& \vec{q}(t) \cdot \vec{\sigma}  \label{piBloch}
\ea
where $\vec{\sigma} = (\sigma_x, \sigma_y, \sigma_z)$ is the vector of Pauli matrices.
Since we need to consider the rotating variables $\tilde{\rho}(t)$ and $\tilde{\mathcal{\pi}}(t)$ introduced in Eqs. \eqref{rhoTilde} and \eqref{piTilde},
we name $\tilde{\vec{a}}(t)$, $\tilde{\vec{q}}(t)$ the associated Bloch vectors.
The PMP conditions allow to find (see Appendix C1) only one extremal solution with non-zero coherences
$\tilde{a}_x^2+ \tilde{a}_y^2 \neq 0$, for which
\ba \label{az1} 
\tilde{a}_z=a_z^{eq}\frac{\left (1+\frac{\beta\epsilon}{\sinh(\beta \epsilon)}\right )}{\left (1-\frac{\beta\epsilon}{\sinh(\beta\epsilon)} \right )},  
\ea 
where
\ba
a_z^{eq}\equiv -\tanh\left (\frac{\beta\epsilon}{2}\right )
\label{aeq}
\ea
is the z-component of the Bloch vector at equilibrium.
However, this solution cannot to be accepted, since it gives $|\tilde{a}_z| > 1 $ for any value of  $\epsilon$, corresponding to non-physical quantum states.
On the other hand, we recover the solution with $\tilde{a}_x(t)=\tilde{a}_y(t)=0$ and $\Lambda(t)=0$, thus exactly reproducing the results of Ref. \cite{Cavina2017bis}. This implies that the most general structure of the optimal coherent protocol for evolving an initial state $\rho(0)$ to a final state $\rho(\tau)$ is the following:
\begin{enumerate}
\item rotate $H(0)$ in a basis in which it is diagonal and commuting with $\rho(0)$;
\item follow the optimal semi-classical process already determined in Ref. \cite{Cavina2017bis} until the state eigenvalues match those of $\rho(\tau)$;
\item perform an instantaneous unitary operation, rotating the state to the desired target $\rho(\tau)$. 
\end{enumerate}
Note that while step 1  is just a quench in the controlled Hamiltonian which does not affect the state of the system, step 3 instead corresponds to a singular perturbation of the Hamiltonian rotating the quantum sate. This means that,  in the ideal situation of achievable unconstrained controls, the only strictly coherent operation on the quantum system is the final unitary rotation. 

For what concerns step 2 of the protocol, corresponding to a regular process lasting for $t\in(0,\tau)$, apparently coherent operations do not help. This means that for initial and final diagonal states of the two-level system, the restriction of the analysis to the set of incoherent protocols only (as performed in Ref. \cite{Cavina2017bis}) was indeed justified.
On the other hand, since this result can be a peculiarity of the Gibbs mixing channel, in the next subsections we will consider two further, different kinds of dynamical evolution.

\subsection{Two-level system in a thermal bosonic bath}
The evolution of a two-level system with Hamiltonian
$H_{\epsilon(t)}= \epsilon(t) \sigma_z/2$ 
in contact with a bosonic heat bath can be described, under phisically reasonable assumptions \cite{Breuer2002}, by the master equation \eqref{dyn} with the following dissipator commonly used in quantum optics
\begin{equation} \label{Otdiss}  
\begin{gathered} 
\mathcal{D}_B[\tilde\rho(t)]= \gamma\biggl \{(1+ N_B) \left[ \sigma_- \tilde \rho(t) \sigma_+ - \frac{1}{2} \{\tilde\rho(t), \sigma_+ \sigma_-\} \right] 
\\ + N_B \left[ \sigma_+ \tilde\rho(t) \sigma_- - \frac{1}{2} \{\tilde\rho(t), \sigma_- \sigma_+\} \right]\biggr \}, \end{gathered} 
\end{equation}
where $ N_B(\epsilon(t)) =  (e^{\beta \epsilon(t) } -1)^{-1} $ is the average
excitation number associated with the energy $\epsilon(t)$, and $\gamma$ is the decoherence rate.
Both dissipators \eqref{master} and \eqref{Otdiss} tend to push the system towards the same equilibrium Gibbs state associated with the instantaneous Hamiltonian, however the thermalization processes are different and therefore we expect different optimal controls.

Before we start our analysis, it is more convenient to express \eqref{Otdiss} in terms of the Bloch coordinates \eqref{rhoBloch}, giving
\ba \label{Otdiss2}
{\cal{D}}_B(\tilde\rho)= \frac{\gamma}{4a_z^{eq}}\left [\vec{\tilde a}\cdot \vec{\sigma} +\left ({\tilde a}_z - 2a_z^{eq}\right )\sigma_z \right ],
\ea
where $a_z^{eq}$ is the same as for the Gibbs mixing channel, Eq. (\ref{aeq}).

As we did in Section IIIA, we first consider a coherent solution of \eqref{dyn} in which $\tilde{a}_z \in [-1,1]$ exists and reads
\begin{gather}  \label{solcor} 
\tilde{a}_z = a_z^{eq}+ \frac{\mu}{\cosh^2(\frac{\beta \epsilon}{2})} \bigg[1 \pm \sqrt{1- \frac{\beta \epsilon}{4 \mu^2} \sinh (\beta \epsilon)}\bigg],
\end{gather}
where we defined $\mu := (\mathcal{K}\beta)/(2 \gamma)$, while the off diagonal terms satisfy
\ba \label{solxy} 
{\tilde a}_x^2 + {\tilde a}_y^2 = 2a_z^{eq}\left (\frac{2 \mathcal{K}}{\gamma\epsilon} -1\right )\big( {\tilde a}_z -a_z^{eq} \big), 
\ea
as proven in Appendix \ref{app:heatbos}.
Equation (\ref{solcor}) in principle describes a set of possible optimal trajectories for ${\tilde a}_z$ as a function of $\epsilon$,
labeled by the conserved quantity $\mathcal{K}$ defined in Eq. \eqref{cons} and by two possible choices of sign (see Appendix \ref{app:heat-examples}).
However, one notes that the right hand side of Eq. (\ref{solxy}) is smaller than zero for all values of $\mathcal{K}$ and $\epsilon$ in the region in which the square root appearing in Eq. (\ref{solcor}) is defined.
We conclude that coherent isothermals are not optimal, similarly to the Gibbs mixing channel.
Then, we look for solutions with no coherence by setting $\tilde{a}_x(t)= \tilde{a}_y(t)=0$. 
Applying the minimum conditions to this case we
obtain the following equation for $\tilde{a}_z$:

\begin{gather} \label{azott} 
{\tilde a}_z = a_z^{(eq)}+  \frac{\mu}{\cosh^2(\frac{\beta\epsilon}{2})}\left [1 \pm
 \sqrt{1 - \frac{\sinh (\beta \epsilon)}{\mu} } \right], 
\end{gather}

where the sign is fixed by the values of $\epsilon$ and $\dot{a}_z$,
as discussed in Appendix \ref{app:heatbos}.
The equation (\ref{azott}) represent the only acceptable regular solution for the 
heat minimization problem when the dynamics is described by the dissipator
(\ref{Otdiss}) and by construction connects only states that 
are diagonal in the energy eigenbasis.
The optimal 
protocol for arbitrary initial and final conditions can be
obtained with the same reasoning of the previous paragraph,
to which is substantially equivalent apart from the intermediate step that is
described by an open evolution of the form
(\ref{azott}) instead of the one derived in \cite{Cavina2017bis}.

\subsection{Two-level system in a thermal fermionic bath \label{sec:ferm}}

Consider now a two level system weakly coupled with a fermionic environment
and suppose that the dynamics is characterized by Eq. (\ref{dyn}), again 
in agreement with the MME approach.
In this case the dissipator reads \cite{Gardiner2004}
\ba 
\label{fermdiss}
\begin{gathered} 
\mathcal{D}_F[\tilde{\rho}(t)] = \gamma\biggl \{(1- N_F) \left[ \sigma_- \tilde \rho(t) \sigma_+ - \frac{1}{2} \{\tilde\rho(t), \sigma_+ \sigma_-\} \right] 
\\ + N_F \left[ \sigma_+ \tilde\rho(t) \sigma_- - \frac{1}{2} \{\tilde\rho(t), \sigma_- \sigma_+\} \right]\biggr \}, 
\end{gathered}
\ea
where $N_F(\epsilon(t)) = (e^{\beta \epsilon(t)} +1)^{-1}$ is the average number of fermionic
excitations in resonance with the system.
Using the Bloch vector parameterization, Eq. (\ref{fermdiss}) becomes
\ba  
\label{endm}
 \mathcal{D}_F(\tilde{\rho}) = - \frac{\gamma}{4} \big[ \vec{\tilde{a}} \cdot
 \vec{\sigma} + (\tilde{a}_z - 2a_z^{eq}) \sigma_z \big],
 \ea
where again $a_z^{eq}$ is given by Eq. (\ref{aeq}).
Thus, the fermionic bath model and
the Gibbs mixing channel considered in Section \ref{Cpm} are strictly related, since the terms in Eq. (\ref{endm}) can be rearranged in order to obtain
 \ba 
 \label{decomp} 
 \mathcal{D}_F(\tilde{\rho}) = \gamma[\eta_{\beta}(t) - \tilde{\rho}(t)] + \frac{\gamma}{4} (\tilde{a}_x \sigma_x + \tilde{a}_y \sigma_y),  
 \ea 
i.e., the evolution in the fermionic scenario is generated by adding a phase damping
component to the Gibbs mixing channel (\ref{master}).
It is easy to show that, since the additional dephasing is independent of the control $\epsilon(t)$, it 
does not play any role in the characterization of the optimal trajectories (see Appendix \ref{app:heatferm}) that, as a consequence, 
are equal to the ones described in Section \ref{Cpm}.
More in details, after showing that the only regular solution of the minimization problem does not involve coherent
operations, it exactly reduces to the one obtained in Ref. \cite{Cavina2017bis}, since the two dissipators 
 (\ref{fermdiss}) and (\ref{master}) act in the same way on the diagonal part of $\tilde{\rho}(t)$.

\section{Application to quantum speed limits, reachable states}
\label{Aqs}

Beyond thermodynamics, the general formalism introduced in Section \ref{variational} can be applied also for determining quantum speed limits and for characterizing the set of {\it reachable states}, {\it i.e.}\ the set of all states reachable via quantum control from a given initial state $\rho(0)$ in a given time interval $\tau$.
In order to minimize the total time required to evolve an open quantum system from an initial state to a final state, we choose the constant functional
\ba
F_{\bold{u}(t)} [{{\rho}}(t)]=\frac{\identity}{\mathrm{Tr}[\identity]},
\ea
in Eq.\ \eqref{generalf} such that the generic cost function $f$ becomes equal to the time length $\tau$ of the process. 

Accordingly, the general pseudo Hamiltonian given in Eq.\ \eqref{HHNew} reduces to
\ba
{\mathcal{H}}(t)&:=& 1 + \Big\langle\tilde{\mathcal{\pi}}(t)\mathcal{L}_{D(t)}[\tilde{\rho}(t)]\Big\rangle  \nonumber \\
&&+ \lambda(t) ( \langle \tilde{\rho}(t)\rangle - 1) + i\langle \Lambda(t) [\tilde{\rho}(t), \tilde{\pi}(t)]\rangle   . \label{HHt} 
\ea 
Then we can apply the PMP conditions listed in Section II to this pseudo Hamiltonian, with the additional constraint (cf. Ref. \cite{Kirk2004}) 
that $\mathcal{K}=0$ in Eq. \eqref{cons}. In other words, the pseudo Hamiltonian computed on shell has to nullify.
We can also compute the equivalent of Eqs. (\ref{4}) that we previously obtained in the heat minimization problem. In the time minimization setting, we obtain the simpler condition
\begin{gather}
\label{4tt}  [\tilde{\mathcal{\pi}}(t), \mathcal{L}_{D(t)}[\tilde{\rho}(t)]] + [\tilde{\rho}(t), \mathcal{L}_{D(t)}^{\dagger}[\tilde{\mathcal{\pi}}(t)]] =0.
\end{gather}

In this way we established a procedure to find quantum speed limits (QSL) \cite{deffner2017} for an open system dynamics with a fully 
controllable Hamiltonian in the 
presence of coherence (for an explicit display of the conditions involved see Appendix D).
It is known that coherence is a resource that can provide a speed boost \cite{marvian2016, Mondal2016} to the evolution of a quantum system, so this kind of 
investigation is interesting {\it per se} since it has a large number of physical applications.
However we want to stress here that the time minimization problem is also interesting from a technical point of view
for the solution of general optimization problems ({\it i.e.} for different functionals, like Eq. (\ref{generalf}) on which this paper is focused)
since it is needed for the characterization of the {\it reachable states} \cite{Kirk2004}.
If there is not enough time to reach the final state, an optimal protocol could not exist, and we can discriminate if this is the case computing the minimum achievable time and comparing it with the total time at disposal.
In the next paragraph we will apply our procedure to the specific case of a two level quantum system, for which the time optimal trajectories 
have been studied in a variety of situations, from the $1/2$-spin particle evolving with Bloch equations \cite{Assemat2013}, to more general 
dissipative maps \cite{Bonnard2009}.
In these physically realizable models the Hamiltonian is not always fully controllable, a paradigmatic example being the optimal control of
a nanomagnetic resonator \cite{Bonnard2012}, in which only the transversal part of the magnetic field is time dependent.
In our model the characterization of the optimal trajectories turns out to be quite simple thanks to the absence of 
constraints on the choice of the external Hamiltonian.

\subsection{Time optimal control of a two-level open system}
Let us consider, for instance, an evolution induced by a master equation of the form (\ref{master})
with the general Hamiltonian \eqref{hameps}
and search for the protocol that allows to go from an initial state $\rho_i$ to a final state $\rho_f$ in the 
minimum time $\tau$.
This analysis will provide also the optimal control strategy for a dynamics induced by Eq.
(\ref{fermdiss}), since we can again exploit the analogy between the two scenarios desribed in sec. \ref{sec:ferm} (see  
Appendix \ref{sec:appferm} for details).
If we call, respectively, $ {\tilde\rho}_{\mathrm{i}} = [\identity + \tilde{a}_z(0) \sigma_z]/2 $ 
and ${\tilde\rho}_{\mathrm{f}} = [\identity + \tilde{a}_z(\tau) \sigma_z]/2$ the diagonalized
versions of the initial and final states, the 
PMP conditions applied to the pseudo Hamiltonian (\ref{HHt}) allow to find 
an optimal trajectory that consists in the following three operations (see Appendix D1):

\begin{enumerate}
\item perform an instantaneous unitary operation that makes $\rho_{\mathrm{i}}$ diagonal in the same basis of the initial Hamiltonian $H(0)$;
\item perform an open evolution of the form (\ref{master}) in which $\epsilon = \pm \infty$, until the state eigenvalues match those of $\rho_{\mathrm{f}}$;
\item perform an instantaneous unitary operation, rotating the state to the desired target $\rho_{\mathrm{f}}$. 
\end{enumerate}

\begin{figure}[t]
\includegraphics[width=0.48 \textwidth]{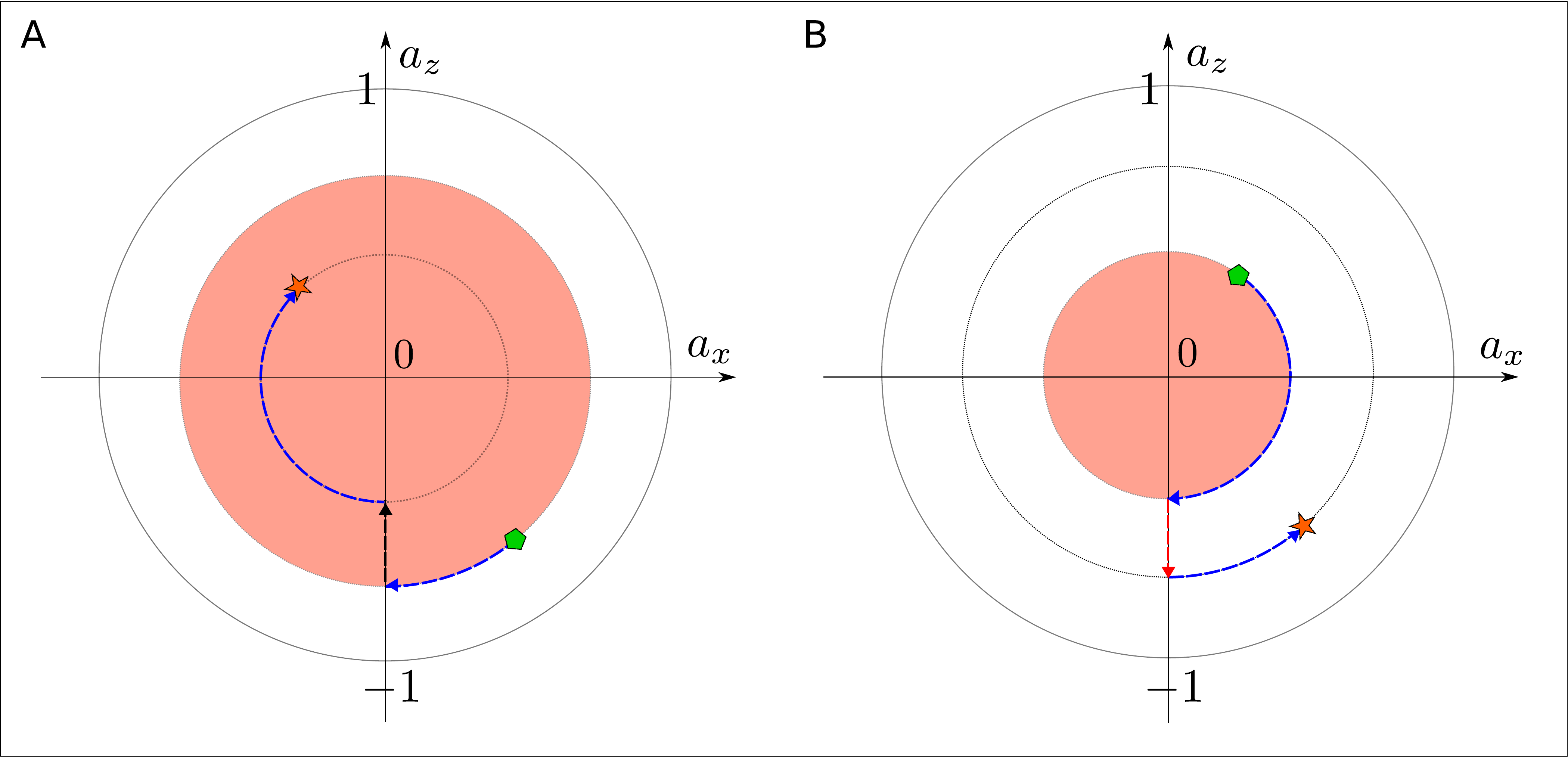}
\caption{Representation in the Bloch sphere of the minimum time trajectories for the Gibbs mixing channel
(\ref{master}) and the bosonic master equation (\ref{Otdiss2}). We suppose that the initial Bloch vectors are in the $x-z$ plane for ease of representation.
A possible optimal trajectory is always composed by two unitary quenches (blue dashed arrows in panel A and B) separated by a semiclassical open 
evolution. The latter depends on the modulus of the initial and final Bloch vectors, as explained
in the main text. 
Panel A: the final state (orange star) has $|\tilde a_z(\tau)|$ smaller than $|\tilde a_z(0)|$ of the initial state
 (green pentagon), so the open evolution (black dashed arrow) occurs respectively with $\epsilon \rightarrow -\infty$ when 
Eq. (\ref{master}) holds and with $\epsilon = 0$ when Eq. (\ref{Otdiss2}) holds.
Panel B: the opposite case, in which $|\tilde a_z(\tau)| > |\tilde a_z(0)|$, the open evolution is such that  $\epsilon \rightarrow \infty$  (red dashed arrow). } \label{fig:2}
 \end{figure}

Note that after step 1 and before step 3 there is a freedom in choosing the sign of $\tilde{a}_z(0)$, which can be switched via a rotation of $\pi$ 
around an axis in the $x-y$ plane. From now on we will always suppose $\tilde{a}_z(0) \leq 0$ and $\tilde{a}_z(\tau) \leq 0$.
Explicitly choosing a diagonal Hamiltonian and $\epsilon= \pm \infty$, Eq. (\ref{master}) generates the following time evolution
\ba \label{expti} 
{\tilde a}_z(t)= {\tilde a}_z(0) e^{-\gamma t} \mp (1-e^{-\gamma t}), 
\label{tott} 
\ea
that allows either an increase or a decrease of $\tilde{a}_z(t)$ depending on the choice of sign:
if $\tilde{a}_z(0) \geq  \tilde{a}_z(\tau) $ we will reach the final configuration only picking $\epsilon \rightarrow \infty$, while the opposite 
choice has to be done otherwise.

\noindent 
The total evolution time $\tau$ is obtained inverting Eq. (\ref{expti})
\ba
\label{scoer} 
\tau = \frac{1}{\gamma} \ln \left[\frac{\tilde{a}_z(0) \pm 1}{\tilde{a}_z(\tau) \pm 1} \right],
\ea
where the sign has to be chosen following the previous prescriptions.
The optimal protocol is summarized in Fig. \ref{fig:2} where we use a convenient representation in terms of the Bloch sphere.
 
We can also apply a similar machinery to a quantum optical evolution of the kind (\ref{Otdiss2}), 
as we did in Appendix D2.
In this framework it is possible to verify that while the structure of the minimum time 
protocol preserves the two quenches and the intermediate open evolution, the 
latter is characterized by different values of the control $\epsilon$.

Indeed if the rotated Bloch coordinates (choosen to be negative) satisfy $\tilde{a}_z(0) \geq \tilde{a}_z(\tau)$ the convenient 
choice turns out to be $\epsilon \rightarrow \infty$ with a total time duration given again by Eq. \eqref{scoer} (with the choice of the plus sign).
Here, however,  in the opposite case $\tilde{a}_z(0) \leq \tilde{a}_z(\tau)$ we have to choose $\epsilon \rightarrow 0^+$ since in this case the optimal time collapses to zero.
This is due to the divergency of the rate for $\epsilon \rightarrow 0^+$ as explained in Appendix D2.
Since in this regime there could be deviations from the Lindblad MME due to the divergency of the coupling strength \cite{Breuer2002, Gorini1976}, in
a more correct optimization procedure the non Markovian corrections have to be taken into account (as discussed, for instance, in \cite{Mukherjee2015}). 

Note that, differently from the optimal relaxation time problem considered in Ref.\ \cite{mukherjee2013}, here we assume that 
the dissipator depends on the system Hamiltonian and therefore it is indirectly affected by the external control.

\section{Conclusions}

We introduced a general formalism suitable for the optimal control of coherent open quantum systems. We first considered the minimization problem associated to a generic linear functional possessing the only property of being covariant with respect to Hamiltonian rotations. Then we applied the general PMP variational techniques to the particular cases of heat minimization and time optimal driving of open quantum systems.
A useful technical contribution of our work is the 
reformulation of the control problem in the instantaneous Hamiltonian eigenframe, that through a time-dependent change of basis allows to introduce an 
effective Hamiltonian term, $\Lambda(t)$, which is responsible for the emergence of quantum coherence between energy eigenstates. This technique allows to significantly simplify the problem leading to many new analytical results and to a characterization of the optimal driving for a two-level system. 
Remarkably, for the three dynamical maps considered in the main text, we are able to show that an optimal coherent regular solution does not exist,
while the only coherent operation is an instantaneous unitary quench performed at the final time.
Other future applications could be the characterization of new quantum speed limits for different kinds of open quantum systems, the optimization of different thermodynamic quantities and the study of thermodynamic cycles. The latter analysis would shed some light on the importance of energy coherence for improving the performances of quantum heat engines.

\section{Acknowledgments}
A.C.'s research was partly funded by Universita' di Pavia (via FQXi Foundation, "Physics of what happens" Program).

\clearpage
\appendix

\section{H-covariance of thermal master equations} \label{app:dissipator}
In order to recover Eq. (\ref{Drot}) we recall that the GKSL dissipator can be written in a standard form \cite{Breuer2002,Lindblad1976},
i.e. in terms of the so called Lindblad operators
 \ba \label{dec} 
 A_{\alpha}(\omega) = \sum_{\omega  = \epsilon - \epsilon'} \Pi(\epsilon) A_{\alpha}  \Pi_{\epsilon'}, 
 \ea
where $\Pi(\epsilon)$ is the projector on the eigenspace with energy $\epsilon$ of the system Hilbert space
while the system Hermitian operators $A_{\alpha}$ appear, along with the associated bath Hermitian operators $B_{\alpha}$, in the decomposition of the interaction Hamiltonian 
$H_I = \sum_{\alpha} A_{\alpha} \otimes B_{\alpha}$.
Directly from Eq. (\ref{dec}) it is easy to check that computing $ A_{\alpha}(\omega) $ with a rotated
system Hamiltonian, is the same as applying the rotation directly to the Lindblad operators.
This is sufficient to prove the following
\ba \label{Drot2} 
\mathcal{D}_{u(t)}[\rho(t)] = U^{\dagger}(t) \mathcal{D}_{D(t)}[U(t)\rho(t)U^{\dagger}(t)] U(t),
 \ea
where we follow the notation introduced in Eq. (\ref{rot}).
Since left and right multiplication by $H(t)$ are clearly $H$-covariant operations, the previous result can be straightforwardly extended to the whole generator of the master equation including the Hamiltonian part. Therefore the property (\ref{Drot}) used in the main text holds and applies to all thermal master equations.

\section{Minimum heat dissipation}  \label{app:heat-general}

Here we focus on the minimization of heat dissipation and we present some details about the calculations leading to the main formulas discussed in the main text.
For instance Eq. (\ref{3}) is obtained from the partial 
derivation of the functional ${\cal{J}}$ with respect to the generator $\Lambda(t)$. The same operation performed with respect to the energy levels $D(t)$ gives
\ba \label{2} 
\Big\langle (\tilde{\mathcal{\pi}}(t) - D(t)) \partial_{i} \mathcal{L}_{D(t)}[\tilde{\rho}(t)]\Big\rangle = \Big\langle  \mathcal{L}_{D(t)}[\tilde{\rho}(t)]
\partial_{i} D(t)  \Big\rangle  , 
\ea
where $\partial_i$ indicates the derivative with respect to the $i$-th diagonal element of $D(t)$.
Note that the previous results hold for the internal region of the space of accessible controls. 
When constraints are introduced a careful inspection for eventual global minima located at the borders of the domain is mandatory.

The condition {\it i)} of the PMP provides the equations of motion (\ref{H1}) that can be
written explicilty as
 \ba  
 \dot{\tilde{\rho}}(t) &=& \mathcal{L}_{D(t)}[\tilde{\rho}(t)] - i[\tilde{\rho}(t), \Lambda(t)],\label{stadyn} 
 \\
 \dot{\tilde{\pi}}(t) &=& \mathcal{L}_{D(t)}^{\dagger} [ D(t) - \tilde{\pi}(t)] - i[\tilde{\pi}(t), \Lambda(t)]-\lambda(t) \identity , \label{cosdyn} 
 \ea
where $\mathcal{L}^{\dagger}$ is the adjoint of the dynamics generator $\mathcal{L}$.
Equation (\ref{4}) is obtained by taking the commutators of $\tilde{\pi}(t)$ and $\tilde{\rho}(t)$ with Eqs. (\ref{stadyn}) and (\ref{cosdyn}), respectively,
adding the two and applying the following identities
\ba 
[\tilde{\pi}(t), [\tilde\rho(t), \Lambda(t)]] = [\tilde{\rho}(t), [\tilde \pi(t), \Lambda(t)]],  
\ea
\ba  [\dot{\tilde \pi}(t),\tilde \rho(t)]+[\tilde \pi(t), \dot{\tilde \rho}(t)]=0, 
\ea
that follow directly from Eq. (\ref{3}) and the Jacobi identity.
Finally the condition {\it iii)} of the PMP gives
\ba \label{energ1} 
\Big\langle (\tilde{\mathcal{\pi}}(t) - D(t)) \mathcal{L}_{D(t)}[\tilde{\rho}(t)]\Big\rangle
 =\mathcal{K}.
 \ea
We note that, in the same way as for Eq. (\ref{4}), this last equation is redundant as it can be 
obtained from the previous conditions (see Refs. \cite{Kirk2004, Cavina2017bis}), and it may be chosen to replace one of the
more cumbersome differential equations \eqref{stadyn}-\eqref{cosdyn}.
 

\section{Control strategies minimizing heat} \label{app:heat-examples}
\noindent

\subsection{Two-level system in a Gibbs mixing channel}

We consider now the heat minimization problem for a two-level system evolving in a Gibbs mixing channel defined by the  master equation with dissipator (\ref{master}).
In the rotating frame, we can express the MME as
 \ba
\label{masterot}{\cal{D}}_G[\tilde{\rho}(t)]&=&\gamma \left [\frac{(\hat I +a_z^{eq} \sigma_z)}{2}-\tilde{\rho}(t)  \right ],
\ea
where $a_z^{eq} $ is the $z$ component of the Bloch vector at equilibrium given by Eq. (\ref{aeq}).
When equations (\ref{hameps}) and (\ref{masterot}) are inserted in the PMP conditions (\ref{3}), (\ref{4}), (\ref{2}) and (\ref{energ1}),  
they give
\begin{gather}
(\tilde{a}_x\tilde{q}_x+\tilde{a}_y\tilde{q}_y)+(\tilde{a}_z -a_z^{eq})\left (\tilde{q}_z -\frac{\epsilon}{2}\right ) =-\frac{\cal{K}}{\gamma}, \label{f1}\\
\frac{\partial a_z^{eq}}{\partial \epsilon}\left (\tilde{q}_z -\frac{\epsilon}{2}\right ) = -\frac{1}{2}(\tilde{a}_z- a_z^{eq}), \label{f2} \\
\vec{\tilde a}\wedge \vec{\tilde q}=\vec{0}, \label{ort} \\ 
a_z^{eq}\tilde{q}_x =-\frac{\epsilon}{2} ~\tilde{a}_x, \label{l1} \\
a_z^{eq}\tilde{q}_y =-\frac{\epsilon}{2}~ \tilde{a}_y . \label{l2}  
\end{gather}
\noindent
Note that in terms of the Bloch vectors $\tilde{\vec{a}}(t)$ and $\tilde{\vec{q}}(t)$,
Eq.\ \eqref{3} becomes the collinearity condition (\ref{ort}), while Eqs.
(\ref{4}) give the last two relations (\ref{l1})-(\ref{l2}).
Moreover,  for the general case of coherent states (i.e., when at least one of the Bloch components $\tilde{a}_x, \tilde{a}_y$ is different from zero), equations \eqref{ort}-\eqref{l2} can be summarized as 
\ba
\vec{\tilde q}=-\frac{\epsilon}{2}\frac{\vec{\tilde a}}{a_z^{eq}}
\label{qgibbs}
\ea
(where we have assumed that $a_z^{eq}\not =0$).
\noindent 

Now we look for a solution with coherence, i.e. for which $ \tilde{a}_x^2 + \tilde{a}_y^2 \neq 0$, so we 
can suppose that at least one between $\tilde{a}_x$, $\tilde{a}_y$ is different from zero.
In particular, substituting the z-component of Eq. \eqref{qgibbs} into
Eq. (\ref{f2}), we can solve for $\tilde{a}_z$ obtaining
\ba\label{az}
\tilde{a}_z
=a_z^{eq}\frac{\left (1+\frac{\partial \ln a_z^{eq}}{\partial \ln \epsilon}\right )}{\left (1-\frac{\partial \ln a_z^{eq}}{\partial \ln \epsilon} \right )} .
\ea
\noindent
Finally, differentiating Eq. \eqref{aeq} we get
\ba
\frac{\partial a_z^{eq}(\epsilon)}{\partial \epsilon}=-\frac{\beta}{2}[1-(a_z^{eq})^2], \label{diffapp}
\ea
and substituting into Eq. (\ref{az}) we eventually obtain Eq.\  \eqref{az1}
of the main text.

Note that, along the steps leading from Eq. (\ref{f2}) to Eq.
(\ref{az}) we divided by the quantity  $ (\partial \ln a^{eq}_z)/(\partial \ln \epsilon ) -1 $, which must be different from zero. However this turns out not to be a physically relevant limitation \cite{note3}.

\subsection{Two-level system in a bosonic thermal bath \label{app:heatbos}}

Let us now consider the minimization problem when the system evolves according to the quantum optics master equation (\ref{Otdiss}), which models the coupling of the system with a bosonic heat bath.
Inserting the dissipator \eqref{Otdiss} into Eqs. (\ref{3}), (\ref{4}), (\ref{2}) and (\ref{energ1}) we obtain 
the minimum conditions for this particular MME
\begin{gather}
(\tilde{a}_x\tilde{q}_x+\tilde{a}_y\tilde{q}_y)+2\left (\tilde{a}_z -a_z^{eq}  \right )\left (\tilde{q}_z -\frac{\epsilon}{2}\right )=2a_z^{eq}\frac{{\cal{K}}}{\gamma}, \label{minott1}
\\
\left [(\tilde{a}_x\tilde{q}_x+\tilde{a}_y\tilde{q}_y)+2 \tilde{a}_z\left (\tilde{q}_z -\frac{\epsilon}{2}\right )\right ] \frac{\partial a_z^{eq}(\epsilon)}{\partial\epsilon}=
-a_z^{eq} \left ({\tilde a}_z -a_z^{eq} \right ),  \label{minott2}
\\
\tilde{\vec{a}}\wedge \tilde{\vec{q}}=\vec{0}, \label{minott3}
\\
\tilde{a}_x\left (\tilde{q}_z -\frac{\epsilon}{2}\right )=a_z^{eq}\tilde{q}_x , \label{minott4}
\\ 
\tilde{a}_y\left (\tilde{q}_z -\frac{\epsilon}{2}\right )=a_z^{eq}\tilde{q}_y . \label{minott5}  \end{gather}
Note that, for the general case of coherent states (i.e., when at least one of the Bloch components $\tilde{a}_x, \tilde{a}_y$ is different from zero), the three equations \eqref{minott3}-\eqref{minott5} imply the more compact condition
\ba
\vec{\tilde q}=\frac{\epsilon}{2}\frac{\vec{\tilde a}}{(\tilde a_z -a_z^{eq})}.
\label{qot}
\ea

Let us first consider the general case including quantum coherences, i.e. let us assume that at least one between the Bloch vector components $\tilde{a}_x$ and $\tilde{a}_y$ is non zero.
Inserting the z-component of Eq. (\ref{qot}) into Eq. (\ref{minott1}) we obtain
\ba \label{coer2} 
\tilde{a}_x \tilde{q}_x+ \tilde{a}_y \tilde{q}_y = a_z^{eq}\frac{(2\mathcal{K}-\gamma \epsilon)}{\gamma}.  
\ea
Then, combining this with Eq. (\ref{minott2}), we find
\begin{gather}
\left [\frac{2\mathcal{K}}{\gamma}+\frac{\epsilon ~a_z^{eq}}{(\tilde{a}_z - a_z^{eq})}\right ]\frac{\partial a_z^{eq}(\epsilon)}{\partial \epsilon}=-(\tilde{a}_z -a_z^{eq}). 
\end{gather}
The latter equation can be simplified introducing the difference between ${\tilde a}_z$ and its equilibrium value (\ref{aeq}), i.e. $\Delta: = \tilde{a}_z -a_z^{eq}$, and we
obtain
\ba  
\Delta^2 + 2 \frac{\mathcal{K}}{\gamma} \frac{\partial{a_z^{eq}(\epsilon)}}{\partial \epsilon} \Delta +\epsilon a_z^{eq}\frac{\partial a_z^{eq}(\epsilon)}{\partial \epsilon}=0.
\ea
The solution of the latter second order equation leads to Eq.\ (\ref{solcor}) of the main text, while Eq. (\ref{solxy}) can be obtained from the previous expression combined with Eqs. (\ref{minott4}) and (\ref{coer2}).

Let us then look for solutions without coherence.
In this case Eqs. (\ref{minott3})-(\ref{minott5}) are trivially verified,
while eliminating $({\tilde q}_z- \epsilon/2)$ from Eqs. \eqref{minott1} and \eqref{minott2}, we are left with
\ba  \label{kott} 
({\tilde a}_z -a_z^{eq})^2 = -2\frac{\mathcal{K}}{\gamma}  {\tilde a}_z\frac{\partial a_z^{eq}(\epsilon)}{\partial \epsilon}.
\ea 
From the last result  and the fact that, from Eq. (\ref{diffapp}), $\partial a_z^{eq}(\epsilon)/\partial \epsilon \leq 0$,
it follows that $\mathcal{K} \leq 0$ if and only if ${\tilde a}_z(t) \leq 0$.
Solving explicitly the Eq. (\ref{kott}) we find an expression for ${\tilde a}_z$ as a function
of the control $\epsilon$, which corresponds to Eq.\ (\ref{azott}) of the main text.

As we can see, from this last expression it is possible to identify two isothermal branches depending on the 
choice of the sign $\pm$.
Since the equation of motion (\ref{Otdiss2}) in the diagonal case reduces to
 $\dot{\tilde a}_z(t) = -\gamma\left [{\tilde a}_z(t) \coth\big(\beta \epsilon/2\big) +1\right ]$, the choice of sign in Eq. (\ref{solcor}) determines the sign of $\dot{\tilde a}_z(t)$.
For instance, if $\epsilon \geq 0$ and $\mathcal{K} \leq 0$ the sign $\pm = - $ characterize an isothermal transformation with
$\dot{\tilde a}_z \leq 0$ in which the heat is released, while the sign $\pm = +$ corresponds to the opposite situation.
Note also that Eq. (\ref{solcor}) is not defined   for negative arguments of the square root, which may happen for $\mathcal{K} \geq 0$ and $\epsilon \geq 0$.

\subsection{Two-level system in a fermionic thermal bath \label{app:heatferm}}

If the dissipative part of the dynamics is regulated by Eq. (\ref{fermdiss}), the PMP conditions 
are only slightly different from those obtained for the Gibbs mixing dissipator (\ref{master}).
Plugging Eq. (\ref{decomp}) into Eqs. (\ref{4}), (\ref{2}) and (\ref{energ1})
we recover Eqs. (\ref{f2}), (\ref{ort}), (C14) and (C15), while Eq. (\ref{f1}) has to be traded with the following
\ba
\label{newferm} 
\frac{1}{2}(\tilde{a}_x\tilde{q}_x+\tilde{a}_y\tilde{q}_y)+(\tilde{a}_z -a_z^{eq})\left (\tilde{q}_z -\frac{\epsilon}{2}\right ) =-\frac{\cal{K}}{\gamma}, \label{fermi1}
\ea
that differs from Eq. (\ref{f1}) only by a multiplicative factor $1/2$ in the first addend on the left hand side.
Following a discussion similar to that for the Gibbs mixing channel, i.e. by substitution of the compact Eq. (C16) (which still holds in the fermionic bath case) into Eq. (C4) and using Eqs. (C2) and (C10), one finds that
\ba
-\left (\frac{\beta\epsilon}{2} \right )\frac{\tanh \left (\frac{\beta\epsilon}{2}\right )}{\cosh^2\left (\frac{\beta\epsilon}{2}\right )}=({\tilde a}_z -a_z^{eq}  )^2,
\ea
which is clearly impossible to be satisfied by any real $\beta\epsilon$.
Thus we conclude that coherent regular solutions are excluded also for the fermionic model. 
It is also easily shown that the solutions without coherence (i.e., with $\tilde{a}_x=\tilde{a}_y=0$) are possible in the fermionic model as well, and they are the same as those presented for the Gibbs mixing channel.
\section{Time optimal control of an open quantum system}\label{app:time}
\noindent

We compute the PMP conditions starting from the pseudo Hamiltonian (\ref{HHt}).
Applying the same procedure followed above for the heat minimization problem,
we obtain in this case the analogue of Eqs. (\ref{3}), (\ref{4}), (\ref{2}) and (\ref{energ1}) in this scenario, i.e.
\begin{gather}
\label{1t} \Big\langle \tilde{\pi}(t) \mathcal{L}_{D(t)}[\tilde{\rho}(t)]\Big\rangle
 =-1,       \\
\label{2t} \Big\langle \tilde{\pi}(t)  \partial_{i} \mathcal{L}_{D_i(t)}[\tilde{\rho}(t)]\Big\rangle = 0, \\ 
\label{3t} [\tilde{\pi}(t), \tilde{\rho}(t)] = 0, \\ \,
\label{4t} [\tilde{\pi}(t), \mathcal{L}_{D(t)}[\tilde{\rho}(t)]] + [\tilde{\rho}(t), \mathcal{L}_{D(t)}^{\dagger}[\tilde{\pi}(t)]] =0.
\end{gather}
Note that, as anticipated in the main text, in Eq. (\ref{1t}) the conserved quantity  has been set to $\mathcal{K}=0$ as required for time minimization problems \cite{Kirk2004}.

\subsection{Two-level system in a Gibbs mixing channel}

Writing Eqs. (\ref{1t}-\ref{4t}) explicitly for the dynamical evolution (\ref{master}) and using Bloch vector coordinates we find
\begin{gather}
(\tilde{a}_x\tilde{q}_x +\tilde{a}_y\tilde{q}_y)+(\tilde{a}_z-a_z^{eq}) \tilde{q}_z=\frac{1}{\gamma}, \label{mas1t} \\
\frac{\partial a_z^{eq}}{\partial \epsilon} \tilde{q}_z=0,  \label{mas2t}  \\
\tilde{\vec{a}} \wedge \tilde{\vec{q}} = 0, \label{mas3t}\\
a_z^{eq} \tilde{q}_x =0,\label{mas4t}\\
a_z^{eq} \tilde{q}_y =0. \label{mas5t} 
\end{gather}

Since the system (\ref{mas1t}-\ref{mas5t}) is sufficient to characterize the optimal trajectory,
let us overview the potential solutions:

\begin{enumerate}
\item We first search for a solution with all coherence terms set equal to zero, i.e.
$\tilde{a}_x(t) = \tilde{a}_y(t) = \tilde{q}_x(t) = \tilde{q}_y(t)=0$ and
$\partial a_z^{eq}/\partial \epsilon=0$, corresponding to the limits $\epsilon=\pm \infty$.
The dynamics of the state following these conditions is 
described by the equation of motion with dissipator (\ref{master}) in the absence of coherence
\ba
\dot{\tilde a}_z= -\gamma ({\tilde a}_z\pm 1),
\ea
subject to Eq. (D5).
The solution of the previous equation for $\tilde{a}_z(t)$ is exactly Eq. (\ref{expti}) of the main text, with corresponding duration time given by Eq. \eqref{scoer}.

\item We then search for a coherent solution such that $\epsilon(t)= a_z^{eq}(t)=0$ and $\tilde{a}_z (t)= \tilde{q}_z (t) = 0$.
In this way the off diagonal elements  $\tilde{a}_x(t)$ and $\tilde{a}_y(t)$ both
relax to zero.  From Eqs. (\ref{mas1t}) and (\ref{mas3t}) we find 
\begin{gather}
|\vec{\tilde a}(t)|=|\vec{\tilde a}(0)|e^{-\gamma t},
\\
\vec{\tilde q}=\frac{\vec{\tilde a}}{\gamma|\vec{\tilde a}|^2}.
\end{gather}
However, this solution turns out to be suboptimal in comparison with the solution without coherences, as one can directly check by computing the total time in the two cases.
Indeed, in the present case the evolution time is 
\ba  
\tau = \frac{1}{\gamma} \ln \left [\frac{|\vec{\tilde a}(0)|}{|\vec{\tilde a}(\tau)|} \right ],  
\ea
which is longer than the time (\ref{scoer}).

\end{enumerate}

We have thus shown that the time optimal open evolution occurs only when the state of the system and the Hamiltonian commute, 
and the complete trajectory is obtained by the composition of the open evolution with two unitary quenches, as
explained in the main text.
However we caution that, since both the basis in which $H$ and $\rho$ are diagonal and the number of unitary quenches are arbitrary, the solution
proposed is locally optimal but not unique.


\subsection{Two-level system in a bosonic thermal bath}

As another example we apply Eqs. (\ref{1t}-\ref{4t}) to the master equation with dissipator (\ref{Otdiss2}) modelling a two-level system in contact with a bosonic heat bath. In this case we get
 \begin{gather} 
 (\vec{\tilde a}\cdot  \vec{\tilde q} + {\tilde a}_z{\tilde q}_z)=-\frac{2a_z^{eq}}{\gamma}(1-\gamma\tilde{q}_z) , \label{condTott1} \\
 (\vec{\tilde a}\cdot  \vec{\tilde q} + {\tilde a}_z{\tilde q}_z) \frac{\partial a_z^{eq}(\epsilon)}{\partial \epsilon} =0, \label{condTott2} \\
\tilde{\vec{a}}\wedge \tilde{\vec{q}} = \vec{0}, \label{condTott3} \\
\left ( \tilde{a}_z -a_z^{eq}\right ) \tilde{q}_x = 0, \label{condTott4} \\
\left ( \tilde{a}_z -a_z^{eq}\right ) \tilde{q}_y = 0. \label{condTott5}
 \end{gather}
From equations (\ref{condTott2}-\ref{condTott5}), it is possible to 
prove that a solution is given by ${\tilde a}_x={\tilde a}_y=0$ and $\epsilon=\infty$ with
\ba\label{tta}
\tilde a_z(t)=[\tilde a_z(0) +1]e^{-\gamma t} -1.
\ea

The optimal trajectory discussed above
is obtained from Eq. (\ref{condTott2}) that provides only 
local stationary points.
However in this dynamical model the equation of motion is non analytical for $\epsilon \rightarrow 0^+$, a point in which the decoherence rate 
diverges, setting the total time to zero.
Applying the condition {\it ii)} of the PMP which states that $\mathcal H(t)$ has to be minimum with respect to the control fields, the limit $\epsilon \rightarrow 0^+$ appears to be the optimal choice.
However, this last solution may not reach all the possible final states as explained in the main text.

\subsection{Two-level system in a fermionic thermal bath \label{sec:appferm}}

If we insert  Eq. (\ref{decomp}) in the minimum condition Eqs. (\ref{1t}), (\ref{2t}), (\ref{3t}) and (\ref{4t})
we recover
 \begin{gather} 
 \frac{1}{2}(\tilde{a}_x\tilde{q}_x +\tilde{a}_y\tilde{q}_y)+(\tilde{a}_z-a_z^{eq}) \tilde{q}_z=\frac{1}{\gamma}, \label{ferm1}\\
\frac{\partial a_z^{eq}}{\partial \epsilon} \tilde{q}_z=0,  \label{ferm2}  \\
\tilde{\vec{a}}\wedge \tilde{\vec{q}} = \vec{0}, \label{ferm3} \\
\left ( \tilde{a}_z -a_z^{eq}\right ) \tilde{q}_x = \frac{\epsilon}{2}\tilde{a}_x, \label{ferm4} \\
\left ( \tilde{a}_z -a_z^{eq}\right ) \tilde{q}_y =  \frac{\epsilon}{2}\tilde{a}_y. \label{ferm5}
 \end{gather}
In particular,  for the general case of coherent states (i.e., when at least one of the Bloch components $\tilde{a}_x, \tilde{a}_y$ is different from zero), 
Eqs. (\ref{ferm3}-\ref{ferm5}) can be substituted by the more compact relation
\ba
\vec{\tilde q}=\frac{\epsilon}{2}\frac{\vec{\tilde a}}{(\tilde a_z -a_z^{eq})}.
\label{ferm6}
\ea
We note that Eq. (\ref{ferm1}) differs from Eq.  (\ref{mas1t}) only by a prefactor $1/2$ on the first addend of the left hand side.
As already discussed when dealing whit minimum-heat trajectories, this factor does not affect the time optimal solution which turns out to be the same as in the Gibbs mixing channel, expressed by Eqs. (38-39).

\section{Explicit equations of motion for the previous examples}
\noindent
For the best convenience of the reader, we write here the set of equations of motion emerging from Eqs. \eqref{H1} for all the examples of optimal control problems considered in this work.

\subsection{Two-level system in a Gibbs mixing channel}
\noindent We parameterize the Hermitian generator of the change os basis as
\ba
\Lambda(t) :=\frac{1}{2}[(\Lambda_0 +\Lambda_3)\identity + 2 (\Lambda_1\sigma_x +\Lambda_2\sigma_y) +(\Lambda_0 -\Lambda_3)\sigma_z  ],
\ea
where $\Lambda_i(t)$ (for $i=0, 1, 2, 3$) are real coefficients.
For the heat minimization problem, the equations of motion for the state Bloch vector are

\ba\label{eqmotionrho}
{\dot {\tilde a}}_x&=&-\gamma {\tilde a}_x +(\Lambda_0-\Lambda_3-\epsilon){\tilde a}_y -2\Lambda_2{\tilde a}_z,
\nonumber \\
{\dot {\tilde a}}_y&=&-\gamma {\tilde a}_y -(\Lambda_0-\Lambda_3-\epsilon){\tilde a}_x +2\Lambda_1{\tilde a}_z,
\nonumber \\
{\dot {\tilde a}}_z&=&-\gamma ({\tilde a}_z-a_z^{eq})+2(\Lambda_2{\tilde a}_x-\Lambda_1{\tilde a}_y),
\ea
while for the costate Bloch vector are
\ba\label{eqmotionpi}
{\dot {\tilde q}}_x&=&\gamma {\tilde q}_x +(\Lambda_0-\Lambda_3-\epsilon){\tilde q}_y -2\Lambda_2{\tilde q}_z,
\nonumber \\
{\dot {\tilde q}}_y&=&\gamma {\tilde q}_y -(\Lambda_0-\Lambda_3-\epsilon){\tilde q}_x +2\Lambda_1{\tilde q}_z,
\nonumber \\
{\dot {\tilde q}}_z&=&\gamma \left ({\tilde q}_z-\frac{\epsilon}{2} \right )+2(\Lambda_2{\tilde q}_x-\Lambda_1{\tilde q}_y).
\ea

\noindent For the time minimization problem, the equations of motion are the same up to the removal of the $\epsilon$ term from the last of the costate equations \eqref{eqmotionpi}.

\subsection{Two-level system in a thermal bosonic bath}
\label{Tls}
\noindent For the heat minimization problem, the equations of motion for the state Bloch vector are

\ba\label{eqmotionrho2}
{\dot {\tilde a}}_x&=&\frac{\gamma}{2a_z^{eq}} {\tilde a}_x +(\Lambda_0-\Lambda_3-\epsilon){\tilde a}_y -2\Lambda_2{\tilde a}_z,
\nonumber \\
{\dot {\tilde a}}_y&=&\frac{\gamma}{2a_z^{eq}} {\tilde a}_y -(\Lambda_0-\Lambda_3-\epsilon){\tilde a}_x +2\Lambda_1{\tilde a}_z,
\nonumber \\
{\dot{\tilde a}}_z&=&\frac{\gamma}{a_z^{eq}} \left ({\tilde a}_z - a_z^{eq}\right )+2(\Lambda_2{\tilde a}_x-\Lambda_1{\tilde a}_y),
\ea
while for the costate Bloch vector are
\ba\label{eqmotionpi2}
{\dot {\tilde q}}_x&=&-\frac{\gamma}{2a_z^{eq}}{\tilde q}_x +(\Lambda_0-\Lambda_3-\epsilon){\tilde q}_y -2\Lambda_2{\tilde q}_z,
\nonumber \\
{\dot {\tilde q}}_y&=&-\frac{\gamma}{2a_z^{eq}}{\tilde  q}_y -(\Lambda_0-\Lambda_3-\epsilon){\tilde q}_x +2\Lambda_1{\tilde q}_z,
\nonumber \\
{\dot {\tilde q}}_z&=&-\frac{\gamma}{a_z^{eq}} \left ({\tilde q}_z-\frac{\epsilon}{2} \right )+2(\Lambda_2{\tilde q}_x-\Lambda_1{\tilde q}_y).
\ea\\

\noindent For the time minimization problem, the equations of motion are the same up to the removal of the $\epsilon$ term from the last of the costate equations \eqref{eqmotionpi2}.
\vspace{1cm}

\subsection{Two-level system in a thermal fermionic bath}

\noindent For the heat minimization problem, the equations of motion for the state Bloch vector are

\ba\label{eqmotionrho3}
{\dot {\tilde a}}_x&=&-\frac{\gamma}{2} {\tilde a}_x +(\Lambda_0-\Lambda_3-\epsilon){\tilde a}_y -2\Lambda_2{\tilde a}_z,
\nonumber \\
{\dot {\tilde a}}_y&=&-\frac{\gamma}{2} {\tilde a}_y -(\Lambda_0-\Lambda_3-\epsilon){\tilde a}_x +2\Lambda_1{\tilde a}_z,
\nonumber \\
{\dot{\tilde a}}_z&=&-\gamma\left ({\tilde a}_z-a_z^{eq}\right )+2(\Lambda_2{\tilde a}_x-\Lambda_1{\tilde a}_y),
\ea
while for the costate Bloch vector are
\ba\label{eqmotionpi3}
{\dot {\tilde q}}_x&=&\frac{\gamma}{2} {\tilde q}_x +(\Lambda_0-\Lambda_3-\epsilon){\tilde q}_y -2\Lambda_2{\tilde q}_z,
\nonumber \\
{\dot {\tilde q}}_y&=&\frac{\gamma}{2}{\tilde  q}_y -(\Lambda_0-\Lambda_3-\epsilon){\tilde q}_x +2\Lambda_1{\tilde q}_z,
\nonumber \\
{\dot {\tilde q}}_z&=&\gamma \left ({\tilde q}_z-\frac{\epsilon}{2} \right )+2(\Lambda_2{\tilde q}_x-\Lambda_1{\tilde q}_y).
\ea\\

\noindent For the time minimization problem, the equations of motion are the same up to the removal of the $\epsilon$ term from the last of the costate equations \eqref{eqmotionpi3}.
\vspace{1cm}

\subsection{Explicitly unravelling the generator $\Lambda$}

It is possible to find a decomposition for $\Lambda$ in the non-rotating frame taking the time derivative of both sides  of Eq. (\ref{rot}),
so that we obtain
\ba \begin{gathered}  \label{lambda}
\dot{D}(t) = \dot{U}(t) H_{u(t)} U^{\dagger}(t) + U(t) \frac{d}{dt} H_{u(t)} U^{\dagger}(t) \\ + U(t) H_{u(t)} \dot{U}^{\dagger}(t).   
\end{gathered} \ea
Sandwiching Eq. \eqref{lambda} between the rotated (fixed) eigenvectors $\bra{\tilde{m}}$ and $\ket{\tilde{n}}$, and 
using $ \ket{n(t)} = U^{\dagger}(t) \ket{\tilde{n}}$, where $\ket{n(t)}$ are the eigenvectors of $H_{\bold u}(t)$ in the non-rotating 
frame, we find
\ba \begin{gathered}   
\delta_{mn} \dot{\epsilon}_n(t) = -[\bracket{\dot{m}(t)}{n(t)} \epsilon_m(t) + \bracket{m(t)}{\dot{n}(t)} \epsilon_n(t)] \\ + 
    \bra{m(t)} \frac{d}{dt} H(t) \ket{n(t)}.
    \end{gathered} \ea

\noindent
Finally, thanks to Eq. (\ref{diffrot}), the off-diagonal elements of $\Lambda$ read
\ba  
\Lambda_{mn}(t) =  i\frac{\bra{m(t)} \frac{d}{dt} H_{\bold u}(t) \ket{n(t)}}{[\epsilon_n(t) - \epsilon_m(t)]}. 
\ea
Thus, from a technical point of view, a direct control of $\Lambda_{nm}(t)$ is equivalent to controlling $ \bra{m(t)} \frac{d}{dt} H_u(t) \ket{n(t)}$,
with the only difference represented by the denominator, that is a regular function if the energy gaps are 
finite (this is consistent with the microscopcal derivation of the Lindblad MME, in particular with the secular 
approximation \cite{Breuer2002}).


\begin{thebibliography}{1}
\bibitem{Nota0} Since the master equation induces a trace preserving evolution, there is some arbitrariness in the definition of the costate. Indeed we can always sum to $\pi(t)$ a term proportional to the identity that fixes a traceless gauge.
\bibitem{notegauge} It is related to the well known global $U(1)$ gauge invariance of quantum mechanics, and thus it is irrelevant both in Eq. (\ref{dyn}) and in Eq. (\ref{heat}).
\bibitem{note3} The choice $ (\partial \ln a^{(eq)}_z)/(\partial \ln \epsilon ) = 1 $ 
satisfies (11) and (\ref{f2})-(\ref{l2}) only when $\vec{\tilde a}=\vec{\tilde q}=0$, which is trivial (i.e., no dynamics 
is allowed for $\tilde\rho$).
\end{thebibliography}
\end{document}